\documentclass[epj]{svjour}
\usepackage{latexsym}
\usepackage{graphics}
\usepackage{epstopdf}
\usepackage{amsmath}
\usepackage{mathtools}
\usepackage{braket}
\usepackage{graphicx}
\usepackage{dcolumn}
\usepackage{bm}
\usepackage{filecontents}
\begin{document}
\title{Lipkin-Meshkov-Glick Model in a Quantum Otto Cycle}
\titlerunning{LMG model as a quantum Otto cycle}
\author{Sel\c{c}uk \c{C}akmak\inst{1,3}, Ferdi Altintas \inst{2,3} \and \"{O}zg\"{u}r E. M\"{u}stecapl{\i}o\u{g}lu\inst{3}
\thanks{\emph{e-mail:} omustecap@ku.edu.tr}%
}                     
%
%
\institute{Department of Physics, Ondokuz May{\i}s University, Samsun 55139, Turkey \and Department of Physics, Abant Izzet Baysal University, Bolu, 14280, Turkey \and Department of Physics, Ko\c{c} University, \.{I}stanbul, Sar{\i}yer 34450, Turkey}
\date{Received: date / Revised version: date}
%
\abstract{
Lipkin-Meshkov-Glick model of two anisotropically interacting spins in a magnetic field is proposed as a working substance of a quantum Otto engine to explore and exploit the anisotropy effects for the optimization of engine operation. 
Three different cases for the adiabatic branches of the cycle have been considered. In the first two cases, either the magnetic field or coupling strength are changed; while in the third case, both the magnetic field and the coupling strength are changed by the same ratio. The system parameters for which the engine can operate similar to or dramatically different from the engines of non-interacting spins or of coupled spins with Ising model or isotropic XY model interactions are determined. In particular, the role of anisotropy to enhance cooperative work, and to optimize maximum work with high efficiency, as well as to operate the engine near the Carnot bound are revealed.
} 

\maketitle
\section{Introduction}\label{intro}
Since the recognition of three level maser as a heat engine~\cite{scovil59}, quantum heat engines (QHEs) have been attracted much interest recently~\cite{scovil59,Quan07,Quan09,Kieu04,Kieu06,Dillen09,Scully03,Hardal15,Huang14,Quan05,Lin03,altintas14,thomas11,zhang08,huang13,feldmann04,feldmann03,huang213,henrich07,zhang07,thomas14,huang12,wu06,ivanchenko15,wang12,he12,huang14plus,wang09,hübner14,azimi14,albayrak13,he2012,alti215,Lutz14,Lutz12,Fialko12,Zhang14,Sothmann12,Quan06,Altintas15}. 
QHEs use quantum matter as their working substance to harvest work from classical or quantum resources through the quantum generalizations of classical thermodynamical cycles, such as Carnot, Otto, Brayton or Diesel cycles~\cite{Quan07,Quan09}. Quantum nature of the working substance, or the resource, can lead to significant advantages. QHEs can extract more work from heat baths relative to classical heat engines~\cite{Kieu04,Kieu06}; they can operate beyond the classical Carnot bound without breaking the second law by exploiting quantum resources; such as entangled~\cite{Dillen09} or quantum coherent heat reservoirs~\cite{Scully03,Hardal15}, or by regenerative steps~\cite{Huang14}. In comparison to non-interacting working substances, such as a two level (qubit)~\cite{Kieu04,Kieu06} or a multilevel atom~\cite{Quan05}, or a simple harmonic oscillator~\cite{Lin03}, QHEs with interacting working substances, in particular coupled spins, are found to be more efficient and capable to harvest more work~\cite{altintas14,thomas11,zhang08,huang13,feldmann04,feldmann03,huang213,henrich07,zhang07,thomas14,huang12,wu06,ivanchenko15,wang12,he12,huang14plus,wang09,hübner14,azimi14,albayrak13,he2012,alti215}.
Physical realizations of QHEs are proposed for a single ion~\cite{Lutz14,Lutz12}, Paul trap~\cite{Huang14}, ultracold atoms~\cite{Fialko12}, optomechanical systems~\cite{Zhang14}, quantum dots~\cite{Sothmann12}, circuit and cavity quantum electrodynamic systems~\cite{Scully03,Quan06,Altintas15}. In implementation of coupled spins however some materials could be used where anisotropy of interactions could be a key parameter to consider. The existing QHE models so far ignore anisotropy effects on the engine efficiency and work output. We would like to address this question in the present contribution. 

As a generic model with anisotropic spin interactions, relevant to several proposed QHE physical systems, we consider the so called Lipkin-Meshkov-Glick (LMG) model~\cite{Lipkin1965,Meshkov1965,Glick1965}. 
Recently, LMG model has received a broad of interest in nuclear physics~\cite{Lipkin1965,Meshkov1965,Glick1965}, magnetic molecules~\cite{Heggie1998}, Bose-Einstein condensates~\cite{Chen09,Cirac1998}, optical cavity quantum electrodynamics~\cite{Morrison2008a,Morrison2008b}, decoheretive systems~\cite{Hamdouni2007,Quan2007a} and in quench dynamics~\cite{Das2006}. The symmetry properties, quantum entanglement and criticality of the LMG model have also been scrutinized, recently~\cite{vidal04,ma2009,ma2011,vidal06,vidal12,vidal10}. Furthermore, LMG model as a small scale quantum thermometer is investigated in Ref.~\cite{Salvatori14}, while the constituent of LMG model given by the single axis twisting model $(\gamma=0)$ as a quantum heat engine and its thermal correlations are analyzed in detail in Ref.~\cite{altintas14}. 

We assume the LMG working substance is subject to a quantum Otto cycle~\cite{Quan07,Quan09}, that consists of two quantum isochoric and adiabatic processes. In the isochoric stages, LMG system is coupled either to a hot bath at temperature $T=T_1$ or to a cold bath at $T=T_2$, and exchanges only heat with the reservoir. LMG system is coupled to a working reservoir during the adiabatic stages, where a net positive work output is expected by changing the parameters in the LMG model. By using their quantum thermodynamical definitions~\cite{Quan07,Quan09}, we calculate the work and efficiency for three different cases of the adiabatic changes. In the first two cases either the external magnetic field or the coupling strength between the spins changes; while in the third case the magnetic field and the coupling strength change simultaneously. According to recent experimental schemes, independent variation of LMG model parameters is possible for example in Bose-Einstein condensates~\cite{Chen09}.

We compare our results with the single qubit Otto engine, which provide a set of benchmark conditions and equations, first investigated by Kieu for a qubit of energy gap $h$~\cite{Kieu04,Kieu06}. For the adiabatic changes between $h_1$ and $h_2$, the efficiency is found to be $\eta=1-h_2/h_1$, which requires the condition $h_1>h_2$ for a qualified Otto cycle. The Otto efficiency is bounded by the Carnot efficiency $\eta_c=1-T_2/T_1$ due to the positive work condition (PWC) $T_1>(h_1/h_2)T_2$. We refer to these conditions as Kieu's conditions for convenience in the subsequent text. The operation of several quantum Otto engines based on a multilevel system~\cite{Quan05}, simple harmonic oscillator~\cite{Quan07}, uncoupled spins~\cite{altintas14,thomas11,alti215} and special coupled spins~\cite{huang14plus} are determined by Kieu's conditions. Present contribution will enquiry to which extent these conditions can be violated for quantum advantages brought by the spin interactions in anisotropic situations.  In addition we compare our results with those similar studies of Heisenberg type interaction coupled spins. These studies showed the interaction enhanced work and efficiency~\cite{altintas14,thomas11,huang13,ivanchenko15,huang14plus,alti215}, and elucidated the possibility of work extraction in the regimes inaccessible by the non-interacting working substances~\cite{altintas14,thomas11}; however they did not take into account anisotropy or LMG type coupling between the spins. Our results reveal unique advantages of anisotropy for enhancing cooperative work output, opening new working regimes, optimization of maximum work at optimal efficiency as well as bring practical advantages by extending the domain of interacting spin working substances to anisotropic systems. 

Our results indicate that with LMG model, one can approach Carnot bound with an Otto cycle. In the first and second cases of the adiabatic changes, we determine the parameters in which Kieu's PWC and efficiency conditions are violated. Indeed, by carefully controlling the system parameters, the LMG Otto cycle can produce positive work with an efficiency near to the classical Carnot efficiency. For the final case, the engine is found to operate at the qubit Otto efficiency. Moreover, it is conjectured that Kieu's PWC is required for the LMG Otto cycle operation. On the other hand, the coupling in that case is found to enhance the work output up to the twelve times of the work produced by a qubit in an Otto cycle.

Finally, we note that there are illuminating studies of the interplay between quantum correlations in the working substance and the work and efficiency of QHEs of coupled spins~\cite{altintas14,zhang08,huang213,zhang07,huang12,wang12,he12,wang09,hübner14,azimi14,albayrak13,he2012}. LMG model is relevant to particle entanglement and with the anisotropy parameter one can tune the system from single axis twisting to two-axes twisting regimes~\cite{Kitagawa93}. In terms of local work and heat concepts~\cite{thomas11,huang13,thomas14,ivanchenko15,huang14plus,alti215}, cooperative effects and the extensitivity breakdown in the global work output~\cite{huang13,huang14plus,alti215} in the LMG system could be examined as well. These intriguing directions could be further explored but beyond the scope of our present contribution.       
\section{\label{subsec:LMG_Model}The Lipkin-Meshkov-Glick model working substance}
LMG model has been originally proposed for a system of $N$ interacting fermions to explore systematically
many-body approximations employed in nuclear physics~\cite{Lipkin1965,Meshkov1965,Glick1965}. 
The model Hamiltonian can be written in terms of pseudo-spin operators as (in a unit system where $\hbar=1$)
\begin{equation} \label{eq:lmg1}
H=-\frac{J}{N}(S_x^2+\gamma S_y^2)-h S_z.
\end{equation}
The model has been used for quantum spin systems as well by taking $S_\alpha = \sum_{i=1}^{N}\sigma_{\alpha}^i/2$ as the total 
spin operators for a set of $N$ spin-$1/2$'s, described by $\sigma_{\alpha}$~($\alpha=x,y,z$) the Pauli matrices. It describes
mutual coupling of spins by Heisenberg $XY$ like interactions, characterized by an interaction strength $J$ and an
anisotropy parameter $\gamma$. In addition the spins are subject to an external transverse magnetic field $h$. 
We take $h\geq 0$ and $-1\leq\gamma\leq 1$, while $J$ can be $J>0$ or $J<0$ in 
ferromagnetic or anti-ferromagnetic systems. $J$ is divided by $N$ in order to take into account Pauli enlargement of volume occupied by the fermions, and the corresponding decrease in the space dependent exchange interaction, as the number of particles is increasing. This normalization keeps the energy per particle finite in the thermodynamical limit.

We consider a two spin system described by the LMG model and take it as the working substance of a quantum Otto engine. Such few spin LMG models are also used in the context of quantum entanglement and geometric phase studies~\cite{eric2010}. The $N=2$ LMG model in terms of Pauli operators becomes
\begin{eqnarray}\label{eq:lmg0}
H=-\frac{J}{4}\left(\sigma_x^1\sigma_x^2+\gamma\sigma_y^1\sigma_y^2\right)-\frac{h}{2}\left(\sigma_z^1+\sigma_z^2\right)
-\frac{J(1+\gamma)}{4}.
\end{eqnarray}
This expression of the model Hamiltonian (\ref{eq:lmg0}) explicitly demonstrates that the coupling in the $x-y$ plane is described by the isotropic XY model when $\gamma=1$. The values in range $-1 \leq \gamma <1$ indicates the anisotropic interaction in the $x-y$ plane, in particular when $\gamma=0$, the coupling is completely polarized along the $x$ direction and is described by the Ising model, while at $\gamma=-1$ the coupling is ferromagnetic in one direction and anti-ferromagnetic in the other.

The eigenvalues and the eigenstates of Eq.~(\ref{eq:lmg0}) in the computational basis $\{\ket{11},\ket{10},\ket{01},\ket{00}\}$ are given by
\begin{eqnarray} \label{eq:lmg3}
E_1&=&0,\hspace{3.2cm} \ket{\psi_1}=\frac{1}{\sqrt{2}}(\ket{10}-\ket{01}),\nonumber\\
E_2&=&-\frac{1}{2}J(1+\gamma),\hspace{1.6cm} \ket{\psi_2}=\frac{1}{\sqrt{2}}(\ket{10}+\ket{01}), \nonumber\\
E_3&=&-\frac{1}{4}[J(1+\gamma)+\kappa], \quad \quad \ket{\psi_3}=(A_-\ket{11}+\ket{00})/\sqrt{1+{\left|A_-\right|^2}},\nonumber\\
E_4&=&-\frac{1}{4}[J(1+\gamma)-\kappa], \quad\quad \ket{\psi_4}=(A_+\ket{11}+\ket{00})/\sqrt{1+{\left|A_+\right|^2}}
\end{eqnarray}
where $\kappa=\sqrt{16h^2+J^2(\gamma-1)^2}$ and $A_{\pm}=(-4h\pm\kappa)/(J(\gamma-1))$.

\section{\label{subsec:QOE}Quantum Otto Cycle}
The working substance with Hamiltonian $H$ and density matrix $\rho$ is manipulated between two heat baths in a quantum Otto cycle which consists of two quantum adiabatic and two quantum isochoric processes. In the adiabatic branches, the system Hamiltonian is changed in an infinitely slow process where quantum adiabatic theorem holds. Therefore, the engine produces zero power output~\cite{ivanchenko15}. Indeed, as shown recently, adiabatic approximation can be safely justified even for quantum systems having degenerate energy spectra~\cite{Rigolin12}. Also, efficient techniques for the adiabatic evolution of the finite size chain LMG model have been given in Refs.~\cite{Caneva08,Caneva09,Solinas08,Zheng15}. In the isochoric stages of the cycle, the thermalization of the working substance with the hot and cold heat baths is assumed for fixed Hamiltonians. The details of the cycle are described as follows.

\textit{Stage 1:} This stage is the quantum isochoric process where the working substance with interaction strength $J=J_1$ and magnetic strength $h=h_1$ attains thermal equilibrium with the hot heat bath at temperature $T=T_1$. The density matrix of the system with Hamiltonian, $H= \sum_{n}E_n\ket{\psi_n}\bra{\psi_n}$, at the end of the stage can be given by the Boltzmann distribution $\rho_1=\sum_{n}p_n\ket{\psi_n}\bra{\psi_n}$ with $p_n=e^{-E_n/T_1}/Z_1$ and $Z_1=\sum_{n}e^{-E_n/T_1}$ (We use a unit system
where $k_B=1$).

\textit{Stage 2:} The working substance is isolated from the hot heat bath and undergoes quantum adiabatic process with changing the magnetic and coupling strengths from $h_1$ to $h_2$ and from $J_1$ to $J_2$, respectively. The system remains in the instantaneous eigenstate of the Hamiltonian, so that the quantum adiabatic theorem holds, i.e., the occupation probabilities of each eigenstates, $p_n$, are maintained during the process~\cite{thomas14}. The Hamiltonian and the density matrix of the system at the end of the stage are, respectively, given as: $H^{'}=\sum_n E_n^{'}\ket{\psi_n^{'}}\bra{\psi_n^{'}}$ and $\rho_1^{'}= \sum_{n} p_n \ket{\psi_n^{'}}\bra{\psi_n^{'}}$.

\textit{Stage 3:} The system is brought into contact to an entropy sink at $T =T_2$ ($T_1>T_2$) which changes the density matrix to $\rho_2=\sum_n p_n^{'} \ket{\psi_n^{'}}\bra{\psi_n^{'}}$ with $p_n^{'}=e^{-E_n^{'}/T_2}/Z_2$, $Z_2=\sum_n e^{-E_n^{'}/T_2}$ at $h=h_2$ and $J=J_2$. The Hamiltonian during the stage is $H^{'}=\sum_n E_n^{'}\ket{\psi_n^{'}}\bra{\psi_n^{'}}$.

\textit{Stage 4:} The working substance undergoes another quantum adiabatic evolution with changing $h_2$ to $h_1$ and $J_2$ to $J_1$ (i.e., $H^{'}$ to $H=\sum_{n}E_n\ket{\psi_n}\bra{\psi_n}$). The density matrix at the end of the stage would be $\rho_2^{'}=\sum_{n}p_n^{'}\ket{\psi_n}\bra{\psi_n}$.

No work is done during the isochoric stages of the cycle. Therefore, the heat exchanges between the heat bath and the working medium in the isochoric stages can be formulated through the change in internal energy $U=Tr[\rho H]$. The heat exchanges in Stage 1 (named $Q_1$) and in Stage 3 (named $Q_2$) can be given as~\cite{Quan05,thomas14}:
\begin{eqnarray} \label{eq:qoc1}
Q_1&=&Tr[H\rho_1]-Tr[H\rho_2^{'}],\nonumber\\ 
&=&\sum_n E_n(p_n-p_n^{'}),\nonumber\\
Q_2&=&Tr[H^{'}\rho_2]-Tr[H^{'}\rho_1^{'}],\nonumber\\
&=&\sum_nE_n^{'}(p_n^{'}-p_n).
\end{eqnarray}
In the adiabatic stages, only work is done. Due to the conservation of energy, the net work done by the QHE can be calculated as:
\begin{eqnarray} \label{eq:qoc2}
W&=&Q_1+Q_2,\nonumber\\
&=&\sum_n (E_n-E_n^{'})(p_n-p_n^{'}),
\end{eqnarray}
where $W>0$ signifies the net work done by the QHE with thermal efficiency $\eta=W/Q_1$ under the constraint $Q_1>-Q_2>0$ due to the second law of thermodynamics.

\section{\label{sec:Results} Results}
To harvest positive work from the LMG model QHE, we envision three scenarios for the adiabatic changes: {\bf (i)} Only the magnetic field changes between two chosen values $(h_1 \rightarrow h_2 \rightarrow h_1)$ at a fixed coupling strength, $J_1=J_2=J$; {\bf (ii)} only the coupling strength changes between two chosen values $(J_1 \rightarrow J_2 \rightarrow J_1)$ at a fixed magnetic field, $h_1=h_2=h$; {\bf (iii)} both the coupling strength and the magnetic field change simultaneously under the condition $r=J_1/h_1=J_2/h_2$, where $r$ is a constant.  

There are some trivial subcases in which the efficiency and the PWC become equivalent to the Kieu's conditions~\cite{Kieu04,Kieu06}.
If $J=0$ in the case (i), the efficiency and PWC becomes $\eta=1-h_2/h_1$ and $T_1>(h_1/h_2)T_2$, respectively. Next to the PWC, one requires $h_1>h_2$ for QHE operation. If $h=0$ in the case (ii), the efficiency becomes $\eta=1-J_2/J_1$, which further requires $J_1>J_2$ for QHE operation. The PWC at $\gamma=0$ and at $\gamma=\pm1$ can be simplified to $T_1>(J_1/J_2)T_2$. In the 
subsequent results, we consider the violations of the Kieu's conditions and  focus on subcases $h_1>h_2$ and $h_1<h_2$ 
in the case (i) and $J_1>J_2$ and $J_1<J_2$ in the case (ii). 

\subsection{\label{sec:case1} Case (i): $h_1 \rightarrow h_2 \rightarrow h_1$ and $J_1=J_2=J$}
\begin{figure}[ht]
	\centering
	\resizebox{0.48\hsize}{!}{\includegraphics{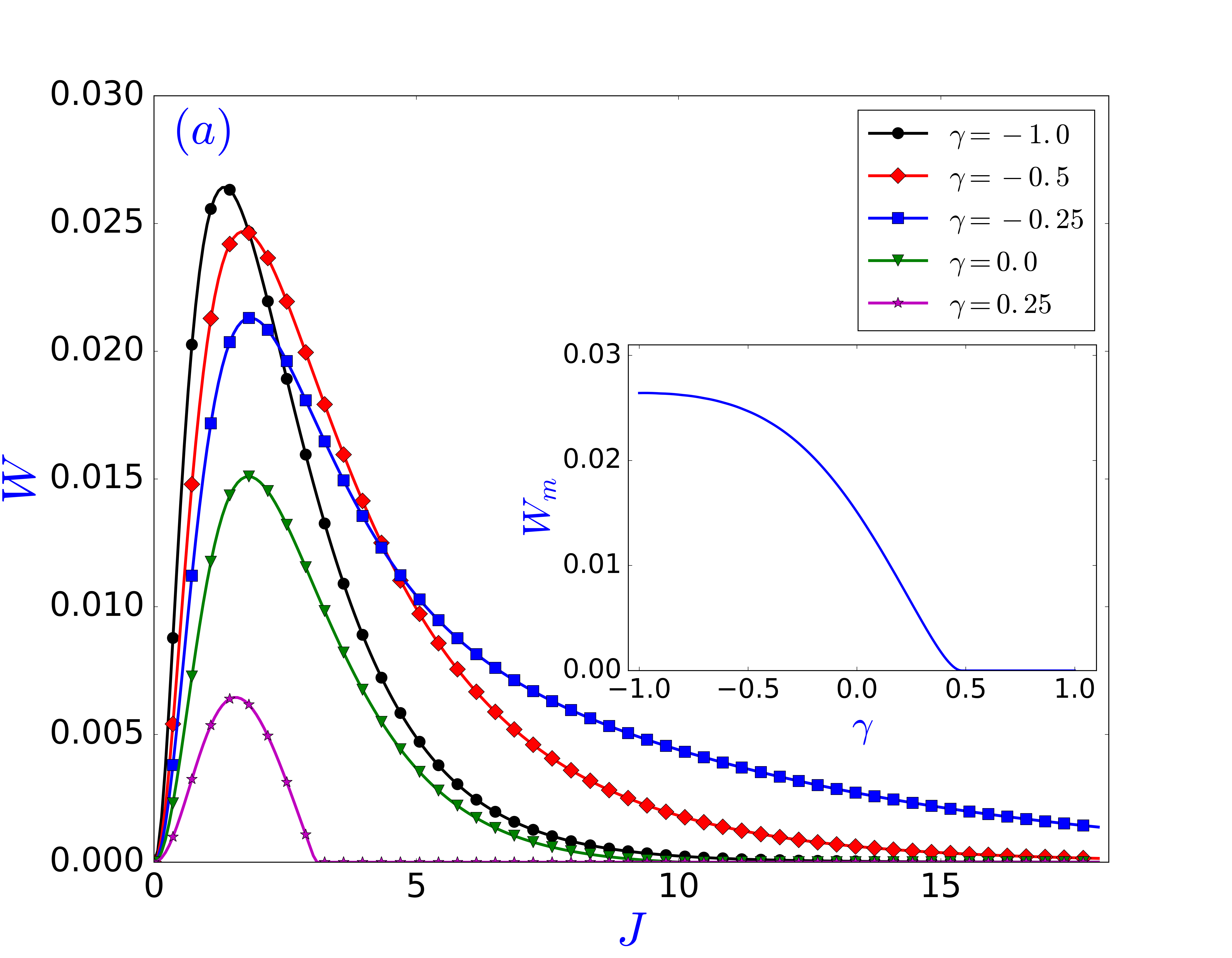}}
	\resizebox{0.48\hsize}{!}{\includegraphics{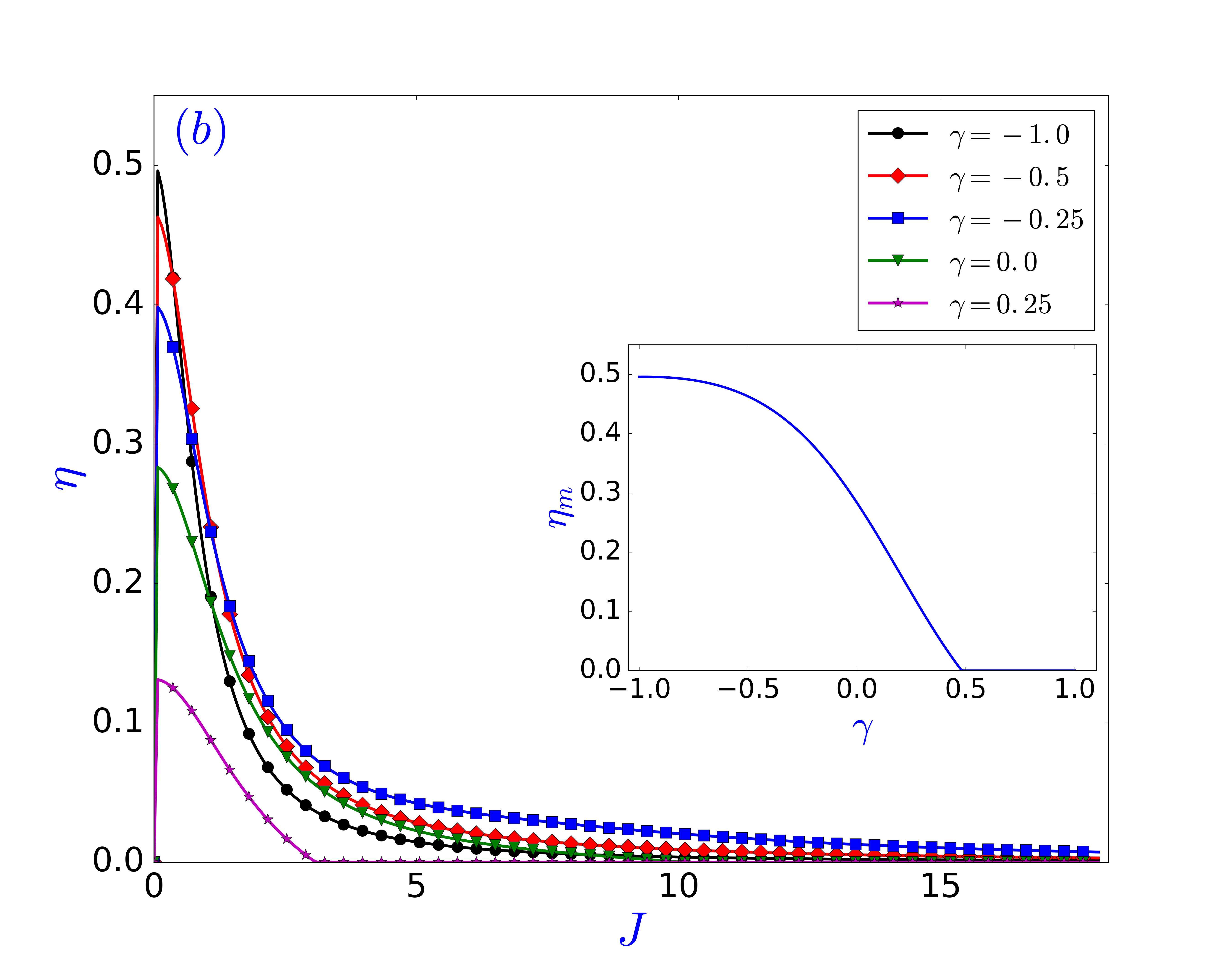}}
	\caption{(Color online.) The work output $W$  (a) and the efficiency $\eta$ (b) as a function of coupling strength $J$ for the parameters $T_1=1.0$, $T_2=0.5$, $h_1=0.50$, $h_2=0.25$ and $\gamma=-1.0,-0.5,-0.25,0.0,0.25$. Insets show $\gamma$ dependence of the maximum work output $W_m$ (a) and the maximum efficiency $\eta_m$ (b) in the full  range of $\gamma$ ($-1\leq \gamma\leq 1$). The Carnot efficiency for the above parameters is $\eta_c=1-T_2/T_1=0.5$. Throughout the paper, we use a unit system where $\hbar=1$ and $k_B =1$.}
\label{fig:qoe0}
\end{figure}

Now we start for the case where only magnetic field is changed in the adiabatic branches and focus on the break down of Kieu's equations in the presence of coupling between spins. In Fig.~\ref{fig:qoe0}, we plot the work output and the thermal efficiency as a function of coupling strength and analyze the effect of coupling strength and anisotropy parameter on the work harvested and its thermal efficiency for the ratios $T_1/T_2=2$ and $h_1/h_2=2$. For the parameters in Fig.~\ref{fig:qoe0}, the Kieu's PWC $T_1>(h_1/h_2)T_2$ at $J=0$ is violated; in the absence of coupling between qubits no positive work output can be obtained from the engine. As shown in Fig.~\ref{fig:qoe0}, positive work can be harvested  at non-zero coupling strengths. It is known that the interacting spins introduce a new PWC on the QHE which depends on $J$ and expands the operational limits of the QHEs relative to uncoupled or special coupled spins~\cite{Quan07,altintas14,thomas11,huang14plus,alti215}.  Here we show that the PWC and the operational regimes are significantly changed by the presence of anisotropy. 

The thermal efficiency is maximum when $J \rightarrow 0$, while the maximum work can be extracted at certain non-zero coupling strengths. Therefore, the engine exhibits an optimal performance at certain coupling strengths in which $W$ is maximum at optimal efficiencies tunable by $\gamma$. Investigations of the effect of $\gamma$ on $W$ and $\eta$ elucidates that $\gamma$ also tunes the allowed $J$-range where the engine harvests work. In the insets of Fig.~\ref{fig:qoe0}, we  report the effect of $\gamma$ on the maximum of the work output and efficiency in the full range of anisotropy parameter, $-1\leq \gamma\leq 1$. The insets show that the engine can operate for a wide range of anisotropies, $-1\leq\gamma<0.45$; while $\eta=0$ for $\gamma>0.45$.  Anisotropy monotonically increases the maximums of the engine efficiency and the work output;  $\gamma=-1$ is the optimal value for the performance of the coupled engine, specifically the engine can produce work at efficiencies near to Carnot bound $(\eta_c=0.5)$ at $\gamma=-1$ (see the inset in Fig.~\ref{fig:qoe0}(b)). The considered thermodynamical quantities are symmetric according to $J\rightarrow -J$, so the results are also the same in the anti-ferromagnetic regime ($J<0$). We should stress here that the flatness of maximum work and efficiency behavior with the anisotropy could bring practical advantage in optimum LMG model QHE implementations.

\begin{figure}[ht]
	\centering
	\resizebox{0.48\hsize}{!}{\includegraphics{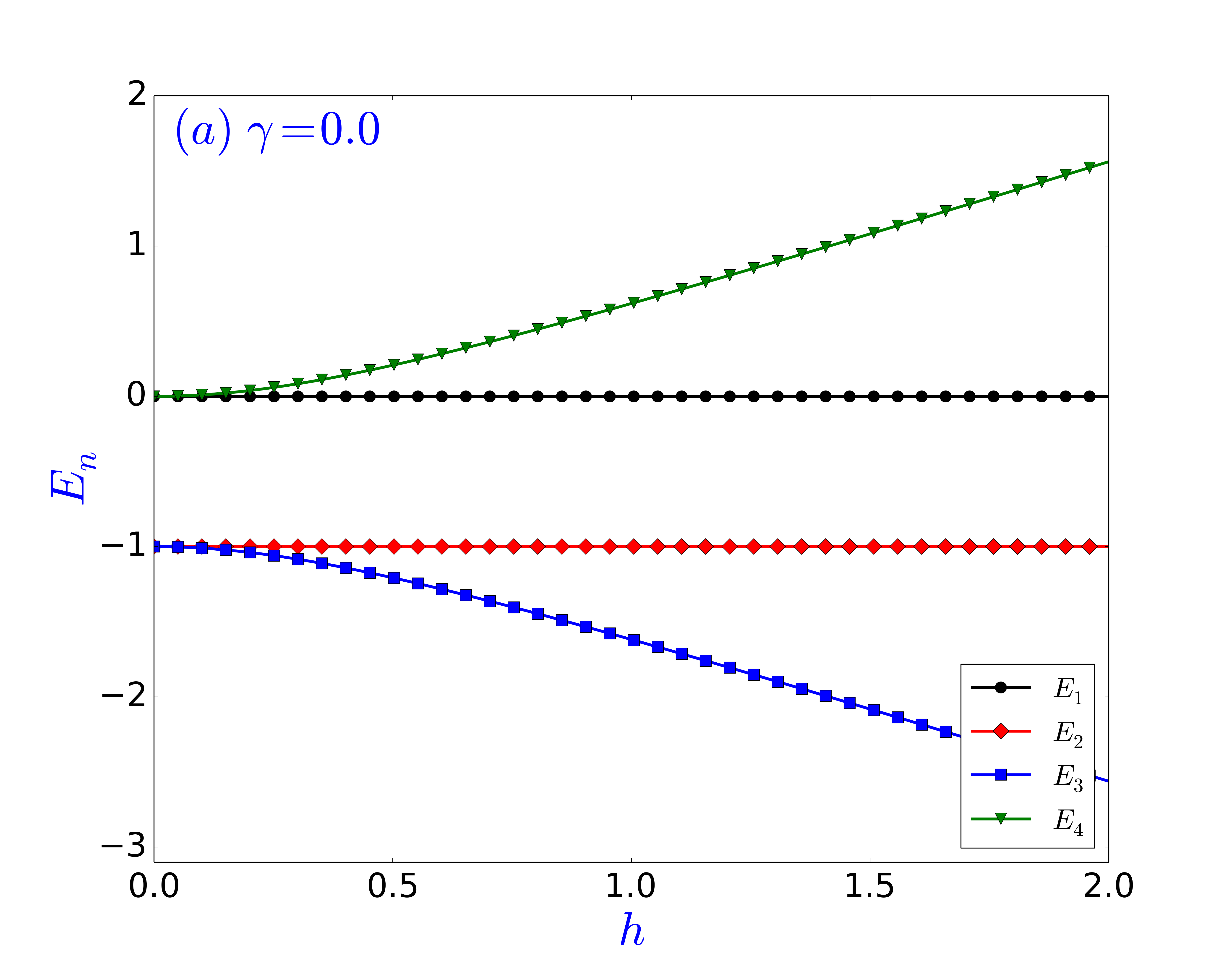}}
	\resizebox{0.48\hsize}{!}{\includegraphics{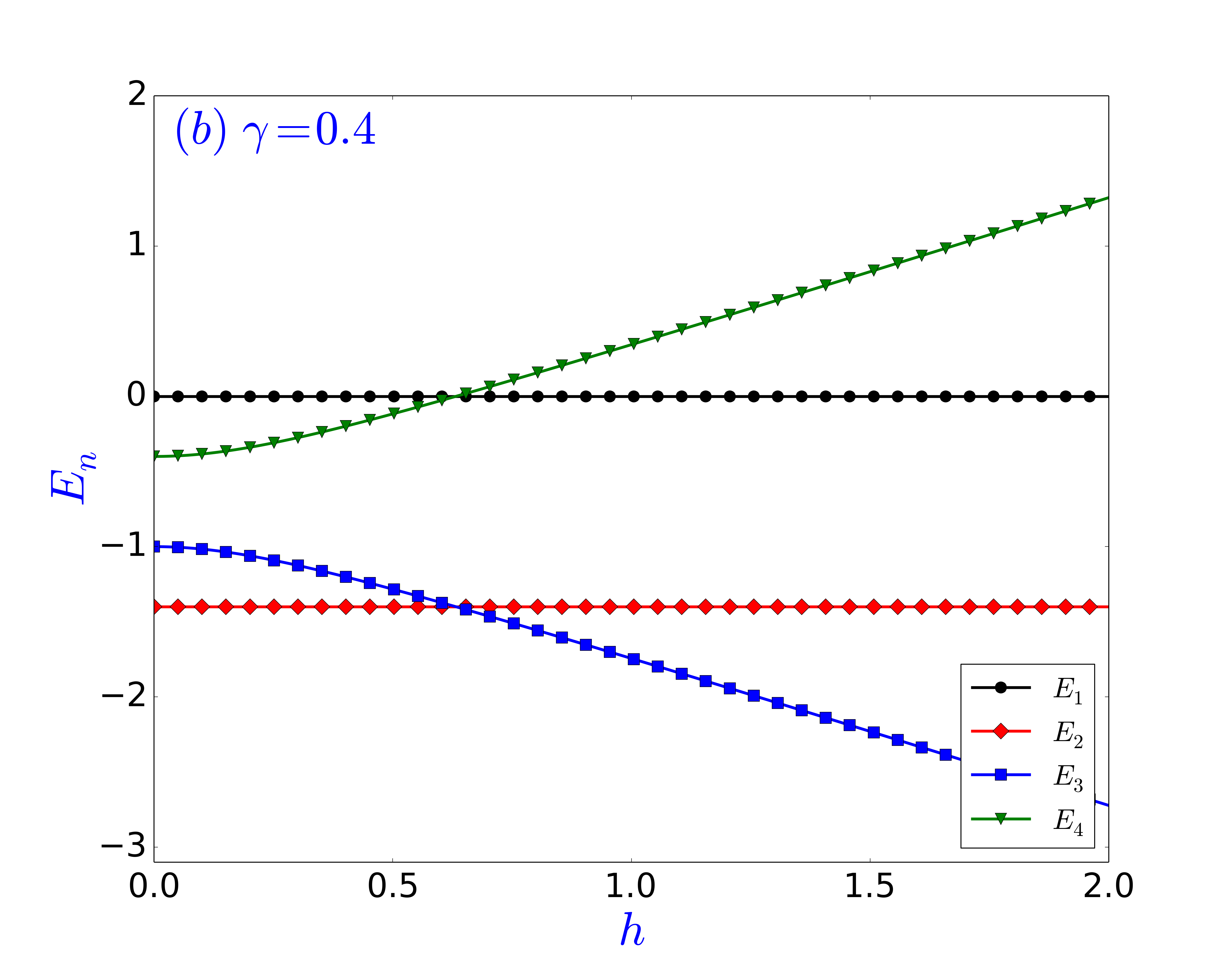}}
	\caption{(Color online.) The energy levels $E_n$ (Eq.~(\ref{eq:lmg3})) as a function of magnetic field $h$ for the values $J=2$ and $\gamma=0$ (a) and $\gamma=0.4$ (b).}
\label{fig:qoe1}
\end{figure}

\begin{figure}[ht]
	\centering
	\resizebox{0.48\hsize}{!}{\includegraphics{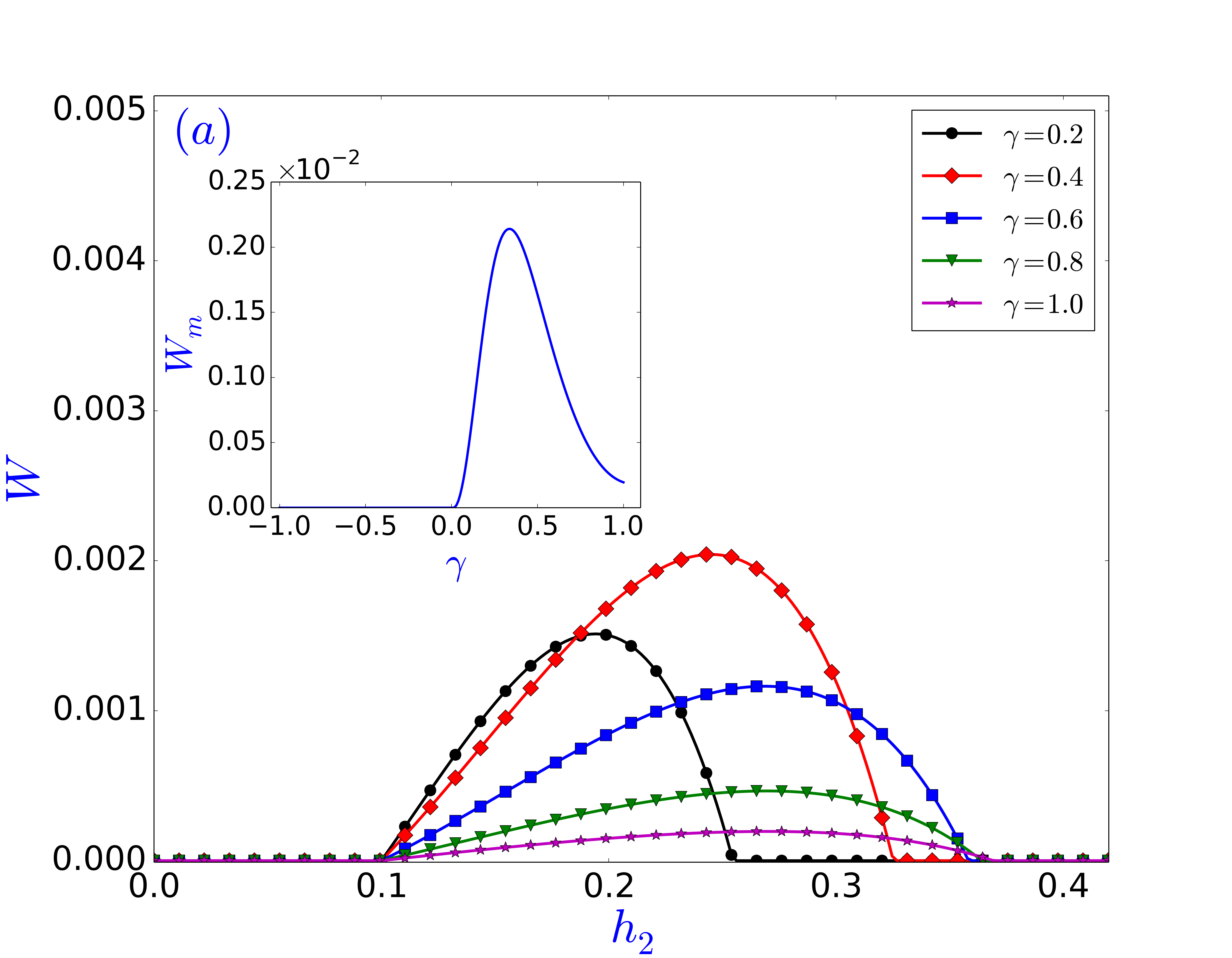}}
	\resizebox{0.48\hsize}{!}{\includegraphics{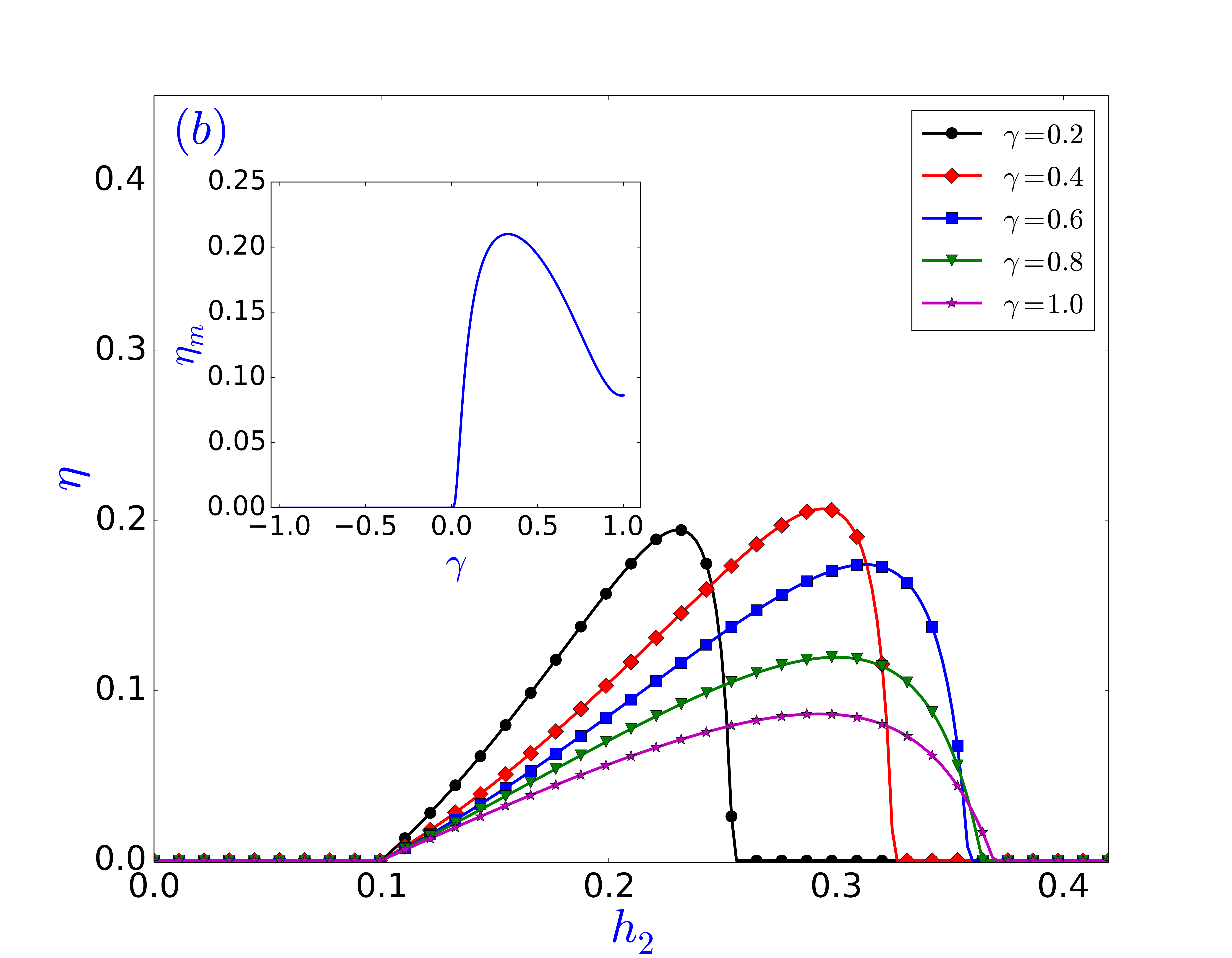}}
	\caption{(Color online.) The work output $W$ (a) and the efficiency $\eta$  (b) as a function of magnetic strength $h_2$ for the parameters $T_1=0.15$, $T_2=0.1$, $h_1=0.1$, $J=2.0$ and $\gamma=0.2,0.4,0.6,0.8,1.0$. Insets show $\gamma$ dependence of the maximum of the work output $W_m$ (a) and the maximum efficiency $\eta_m$ (b) for the heat engine operation in the regions $h_2>h_1$. The Carnot efficiency for the above parameters is $\eta_c=1-T_2/T_1=1/3$.}
\label{fig:qoe2}
\end{figure}

Now we focus on the possibility of work extraction from the coupled engine when it operates under the conditions $h_1<h_2$. An uncoupled engine ($J=0$) with efficiency $\eta=1-h_2/h_1$ cannot operate in this regime, as the energy gaps are shrinking in the second adiabatic stage which are exposing to an adiabatic expansion process and is inconsistent with the behavior expected under an adiabatic compression~\cite{Quan07}. For non-zero $J$, the energy levels at $\gamma=0$ are shown in Fig.~\ref{fig:qoe1}(a). Clearly, the energy gaps becomes wider with larger $h$, as in the case of uncoupled spins. Therefore, it would also not be possible to extract positive work by the coupled engine if $h_1<h_2$ at $\gamma=0$~\cite{altintas14}. However non-zero anisotropy makes a dramatic change and introduces level crossings, as shown in Fig.~\ref{fig:qoe1}(b), which allows for work harvesting in a new regime $h_1<h_2$, inaccessible to uncoupled spins as well as coupled spins at $\gamma=0$. The results are given in Fig.~\ref{fig:qoe2} where the work (Fig.~\ref{fig:qoe2}(a)) and the thermal efficiency (Fig.~\ref{fig:qoe2}(b)) are plotted as a function of $h_2$ and we set $h_1=0.1$. According to Fig.~\ref{fig:qoe2} positive work emerges just after $h_2>h_1$ when $\gamma>0$. Compared to the classical Carnot efficiency $\eta_c=1/3$, the engine can produce work with significantly high efficiency $\eta_m \approx 0.22$. At the low temperatures used in Fig.~\ref{fig:qoe2}, only the lower energy levels are populated and can dominantly contribute to the work extraction. Therefore, the first energy gap, which is exposed an adiabatic compression rather than expansion in the second adiabatic stage for the changes $h_1<h_2$, is responsible for the work extraction in Fig.~\ref{fig:qoe2}. Investigating the $\gamma$ ($>0$) effect on the performance of the engine demonstrates that anisotropy shrinks the allowed $h_2$ range of the possible work extraction. $W$ and $\eta$ are maximums at certain $h_2$ values which are controlled by $\gamma$. The role of $\gamma$ on the maximum of $W$ and $\eta$ are analyzed in detail in the insets of Fig.~\ref{fig:qoe2}. Contrary to the $\gamma$ dependence of $W_m$ and $\eta_m$ in Fig.~\ref{fig:qoe0}, they do not have any monotonic dependence on $\gamma$; $\gamma \approx 0.4$ is the optimal value for the heat engine operations, while $\eta=0$ in the regions $h_1<h_2$ when $\gamma<0$.

\subsection{\label{sec:case2} Case (ii): $J_1 \rightarrow J_2 \rightarrow J_1$ and $h_1=h_2=h$} 
\begin{figure}[ht]
	\centering
	\resizebox{0.48\hsize}{!}{\includegraphics{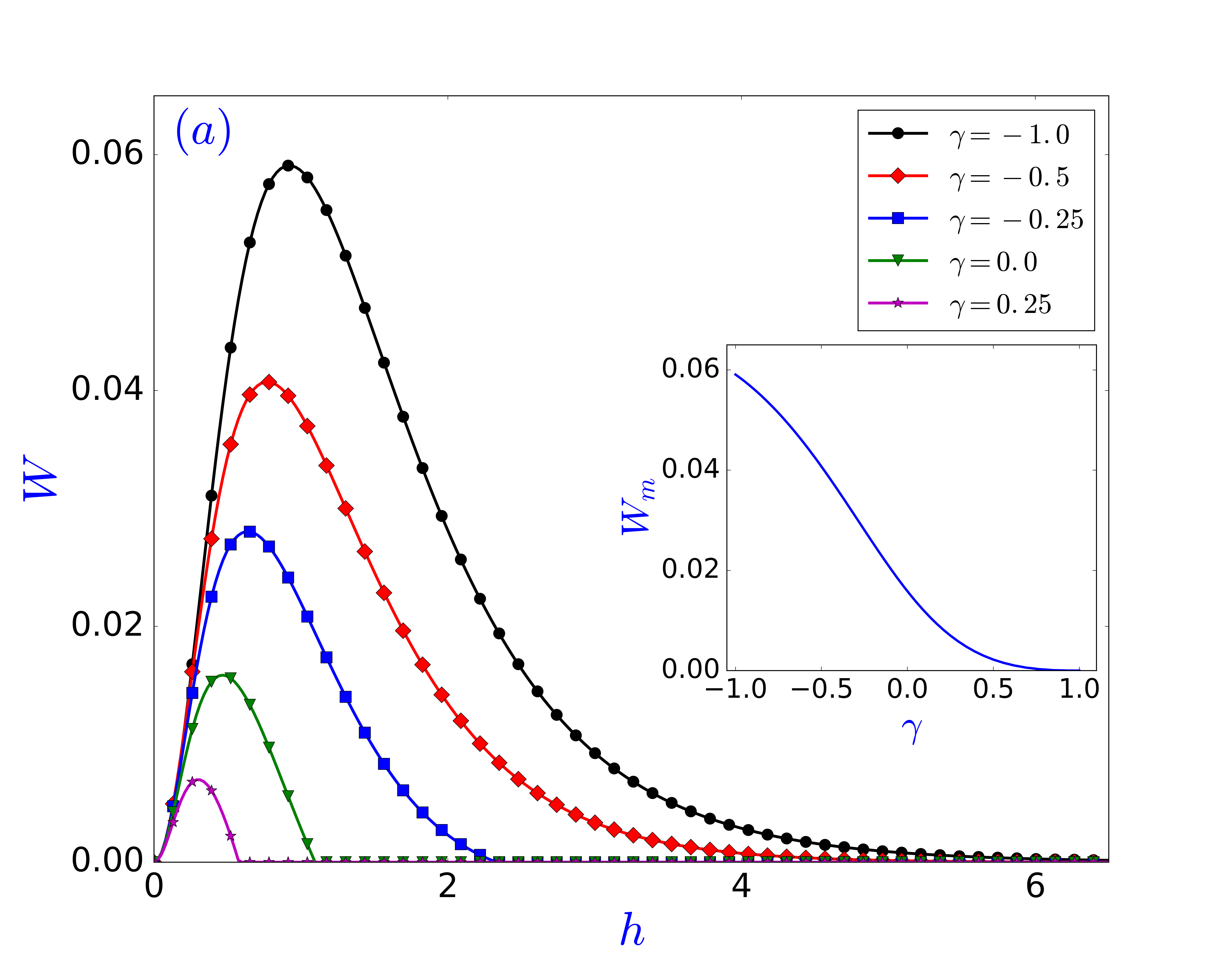}}
	\resizebox{0.48\hsize}{!}{\includegraphics{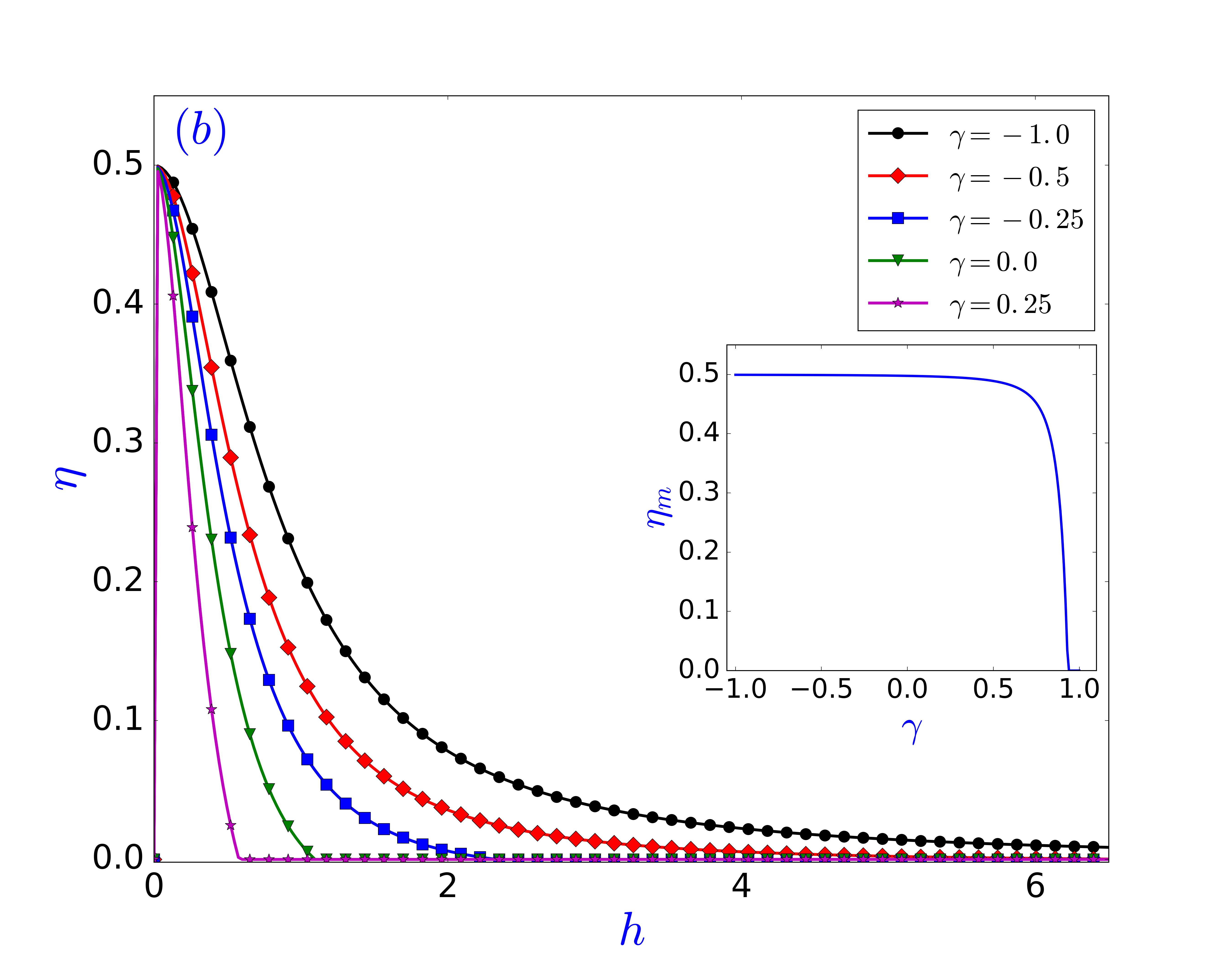}}
	\caption{(Color online.) The work output $W$ (a) and the efficiency $\eta$  (b) as a function of magnetic field $h$ for the parameters $T_1=1.0$, $T_2=0.5$, $J_1=2.0$, $J_2=1.0$ and $\gamma=-1.0,-0.5,-0.25,0.0,0.25$. Insets show $\gamma$ dependence of the maximum of the work output $W_m$ (a) and the efficiency $\eta_m$ (b) in the full range of $\gamma$ ($-1\leq \gamma\leq 1$). The Carnot efficiency for the above parameters is $\eta_c=1-T_2/T_1=0.5$.}
	\label{fig:qoe3}
\end{figure}

Now we focus on the case where the adiabatic branches involve altering only the interaction strength $(J_1 \rightarrow J_2 \rightarrow J_1)$ at a fixed magnetic field, $h_1=h_2=h$. In Fig.~\ref{fig:qoe3}, we plot the work obtained from the QHE and its thermal efficiency as a function of magnetic field strength $h$ for the adiabatic changes in the ferromagnetic region with the ratios $T_1/T_2=2$ and $J_1/J_2=2$. The corresponding thermodynamical quantities are invariant under the simultaneous replacement $J_1\rightarrow -J_1$ and $J_2\rightarrow -J_2$. Therefore, the results are also the same for the adiabatic changes in the anti-ferromagnetic regions. At $h=0$ we have the Kieu's like equations, the efficiency $\eta=1-J_2/J_1$ and the PWC $T_1>(J_1/J_2)T_2$ at $\gamma=0$ and $\gamma=\pm 1$. While there are no similar simple expressions at arbitrary $\gamma$, Fig.~\ref{fig:qoe3} shows that PWC for an arbitrary $\gamma$ is consistent with the usual PWC at $h=0$, where $W=0$. At non-zero $h$, engine can produce useful work with an appreciably high efficiency close to the Carnot bound. The anisotropy and magnetic field dependence of the work output and the efficiency in Fig.~\ref{fig:qoe3} show similar qualitative behavior for the case (i) as investigated in Fig.~\ref{fig:qoe0}; the work output is maximum at certain magnetic strengths, while efficiency is a monotonously decreasing function of $h$ and approaches its maximum as $h \rightarrow 0$. The $\gamma$ dependence of the maximum of the work output and the efficiency in the full range of anisotropy parameter is shown in the insets of Fig.~\ref{fig:qoe3}. Analyzing $\gamma$ effects on the performance of the QHE signifies that anisotropy widens the allowed $h$ range of the possible work extraction and also enhances the maximum work output. As shown in the inset of Fig.~\ref{fig:qoe3}(a), $W=0$ when $\gamma>0.97$.  On the other hand, the inset in Fig.~\ref{fig:qoe3}(b) elucidates that the engine can produce useful work with an efficiency very close to the classical Carnot efficiency in almost all the range of $\gamma$ $(-1 \leq \gamma < 0.97)$. 

\begin{figure}[ht]
	\centering
	\resizebox{0.48\hsize}{!}{\includegraphics{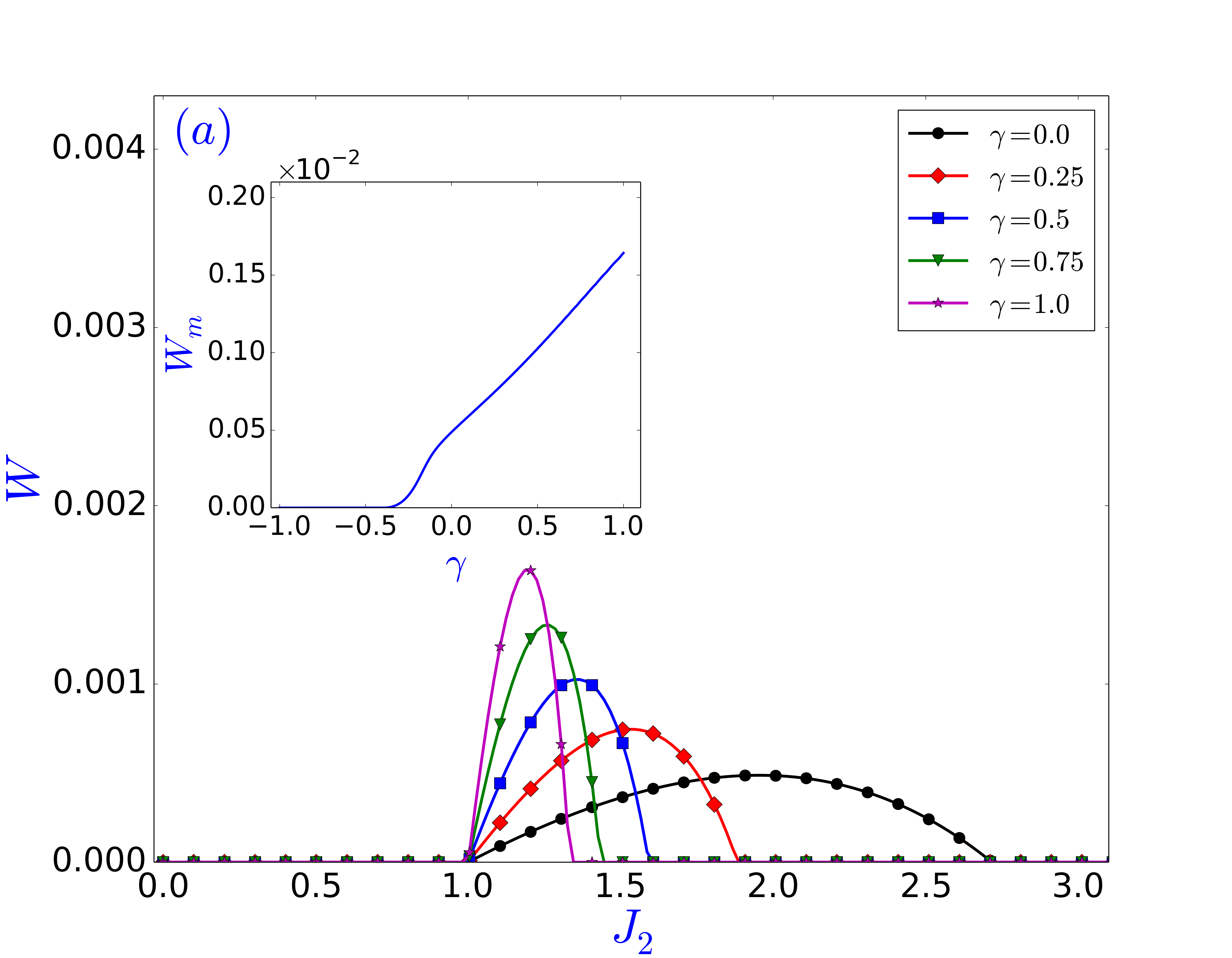}}
	\resizebox{0.48\hsize}{!}{\includegraphics{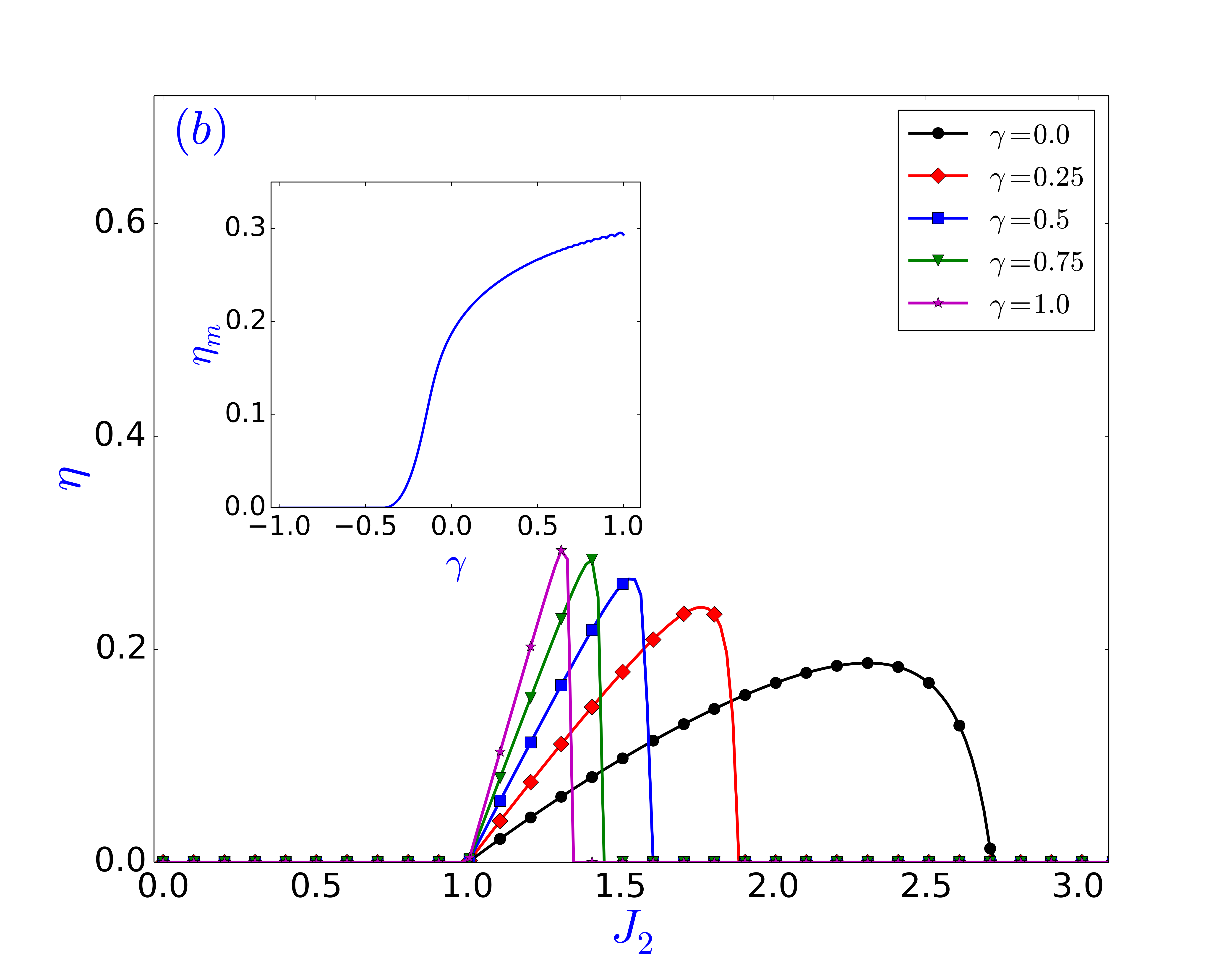}}
	\caption{(Color online.) The work output (a) and the efficiency (b) as a function of the coupling strength $J_2$ for the parameters $T_1=0.15$, $T_2=0.1$, $h=1.0$, $J_1=1.0$ and $\gamma=0.0,0.25,0.50,0.75,1.0$. Insets show $\gamma$ dependence of the maximum of the work output (a) and the efficiency (b) for the heat engine operation in the regions $J_2>J_1$. Note that the Carnot efficiency for the above parameters is $\eta_c=1-T_2/T_1=1/3$.}
	\label{fig:qoe4}
\end{figure}

\begin{figure}[ht]
	\centering
	\resizebox{0.48\hsize}{!}{\includegraphics{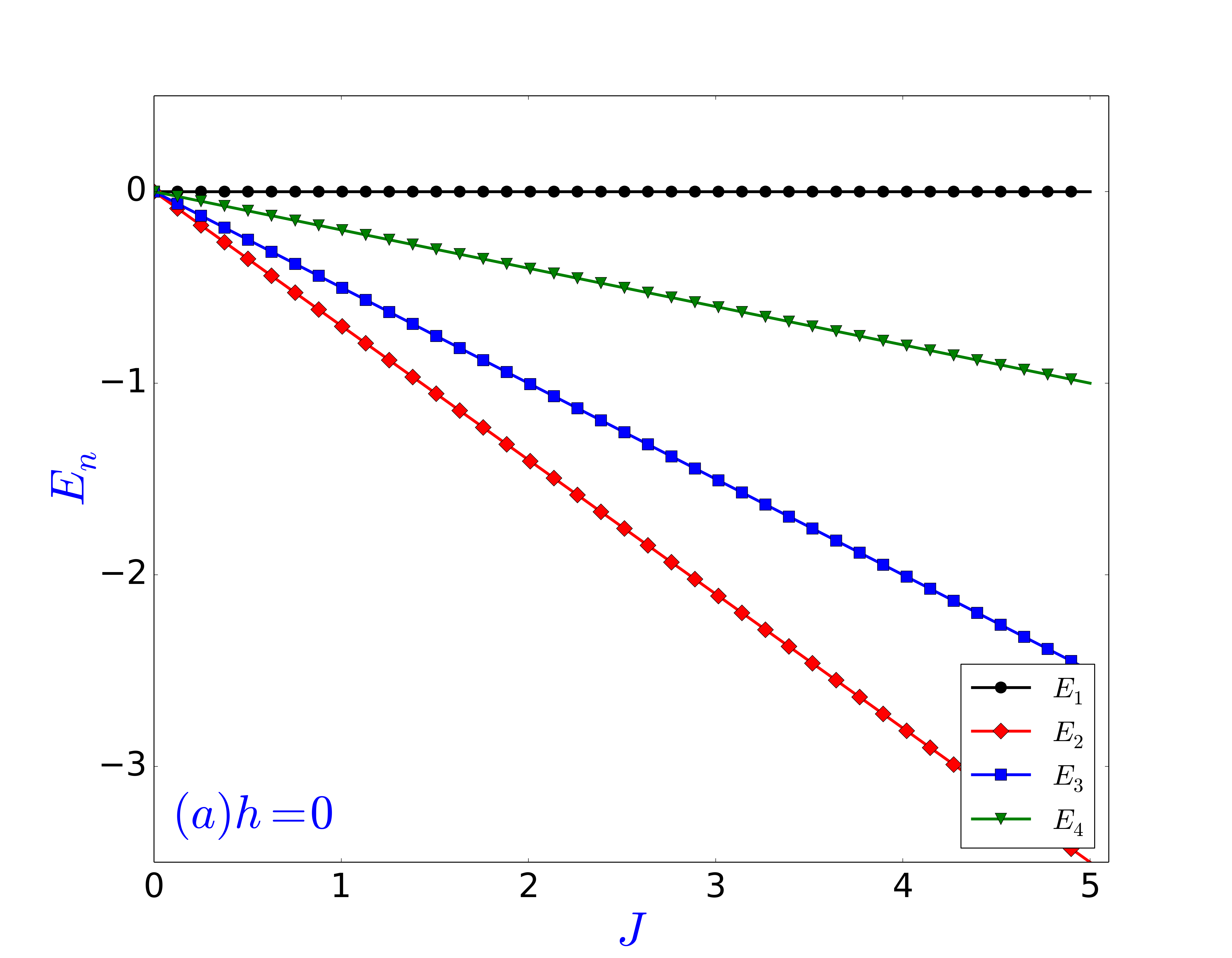}}
	\resizebox{0.48\hsize}{!}{\includegraphics{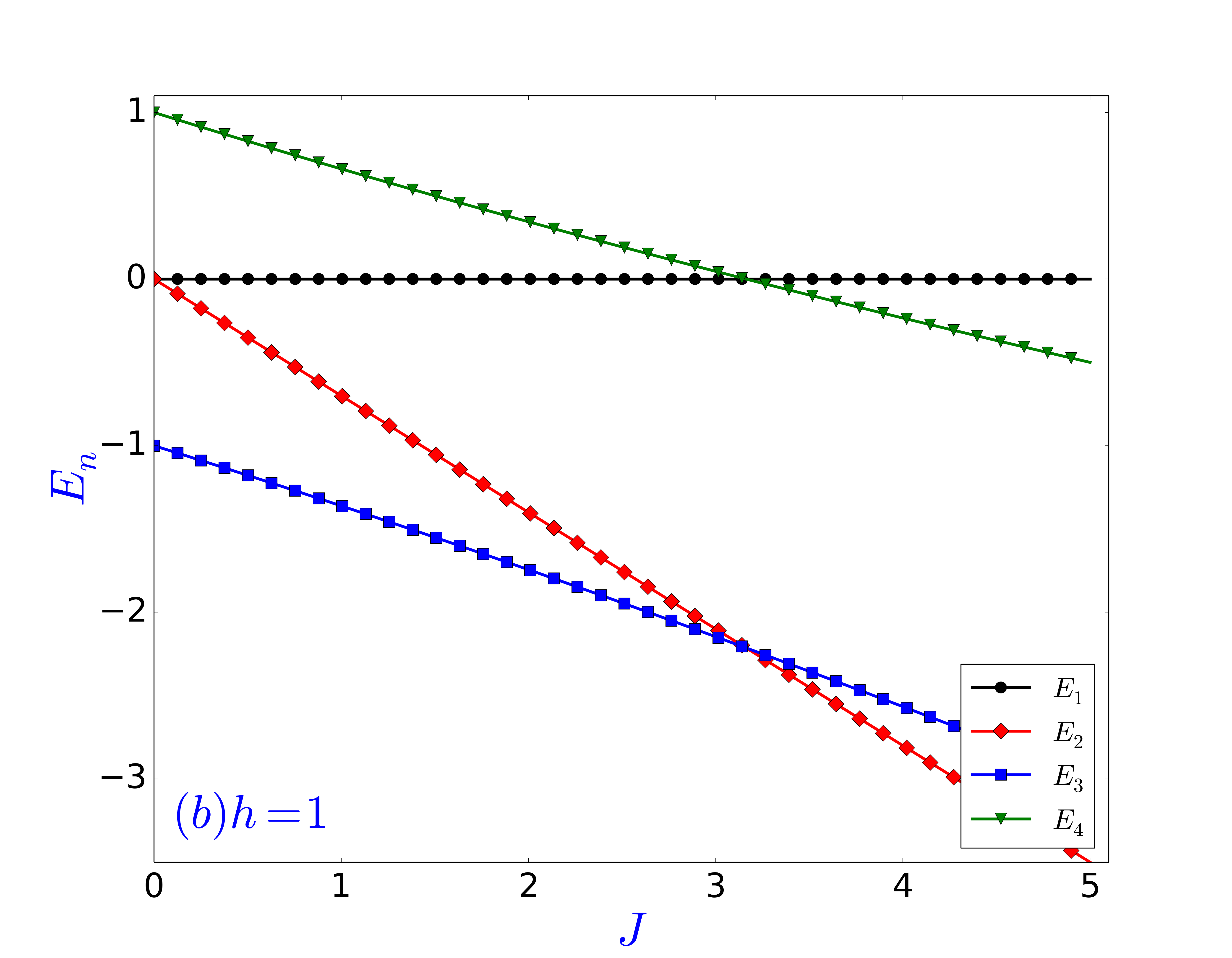}}
	\caption{(Color online.) The distribution of energy levels $E_n$ (Eq.~(\ref{eq:lmg3})) as a function of coupling strength $J$ at $\gamma=0.4$, $h=0$ (a) and $h=1$ (b).}
	\label{fig:qoe5}
\end{figure}

In Fig.~\ref{fig:qoe4}, we plot the work output and the thermal efficiency as a function of $J_2$ for the parameters $T_1=0.15$, $T_2=0.1$, $h=1$ and $J_1=1$ to show the capability of the coupled QHE to produce work even when $J_2>J_1$. For $h=0$, the efficiency is $\eta=1-J_2/J_1$ which requires $J_1>J_2$ for a QHE. The energy levels at $h=0$ (Fig.~\ref{fig:qoe5}(a)) show that $J_1>J_2$ is the condition for the adiabatic compression behaviour in the second adiabatic stage. In addition, when there is a magnetic field, there are level crossings and the first energy gap, which is dominant in the work extraction for the considered temperature parameters, is now able to expand in the second adiabatic stage and is exposing an adiabatic compression process~\cite{Quan07} so that QHE can produce work even for $J_2>J_1$. Compared to the classical Carnot efficiency ($\eta_c=1/3$), the engine can  produce work in the regions $J_2>J_1$ with an appreciably high efficiency; as shown in the inset of Fig.~\ref{fig:qoe4}(b) $\eta_m\approx 0.3$ at $\gamma=1$. Investigating the role of $\gamma$ on the work output and efficiency show that approaching to isotropic coupling ($\gamma \rightarrow 1$) increases the maximums of $\eta$ and $W$, while reducing the $J_2$ range in which the coupled system operates as a QHE; when $\gamma<-0.4$, $\eta$ is zero in the regions $J_2>J_1$. In comparison to Ising model ($\gamma = 0$)~\cite{altintas14}, the anisotropy $\gamma>0$ leads to significant work output and high efficiency. 

\subsection{\label{sec:case3} Case (iii): $J_1/h_1=J_2/h_2=r$} 
In this section, we consider a special coupled four level quantum Otto engine where the coupling strength $J$ and the magnetic field $h$ are both varied in the adiabatic stages obeying the proportionality $J_1/h_1=J_2/h_2=r$~\cite{huang213,huang14plus}. In the adiabatic branches, the magnetic field and the coupling constant are varied simultaneously with the same ratio $r=J/h$; here $r$ can be seen as the relative coupling strength and it is fixed during the cycle. Compared to the previously discussed cases where either $h$ or $J$ is changed during the adiabatic stages, the realization of the simultaneous change require more demanding and sophisticated tasks from the experimental point of view. 

We first focus on the thermal efficiency and the PWC of the cycle. By putting $J_1=rh_1$ and $J_2=rh_2$, one can simplify the engine efficiency for an arbitrary $\gamma$ as:
\begin{equation} \label{eq:case3e0}
\eta = 1-\frac{h_2}{h_1},
\end{equation}
which is equal to the uncoupled engine efficiency as well as the qubit Otto engine efficiency and assigns the strict condition $h_1>h_2$ for a QHE. Remark that the special coupled Otto engine efficiency is independent of all the system parameters, except the magnetic fields $h_i$ $(i=1,2)$.  It can be conjectured that the engine also requires the condition $T_1>(h_1/h_2)T_2$ to harvest positive work. Therefore, the operation conditions of the special coupled Otto cycle are found to be determined by the Kieu's equations. In fact, the LMG Otto cycle with the adiabatic changes $J_1/h_1=J_2/h_2=r$ is equivalent to the special multilevel Otto cycle recently proposed by Quan {\it et al.,} in Ref.~\cite{Quan07}. In this special engine all the energy gaps in the quantum adiabatic stages are changed by the same ratio obeying $E_n-E_m=\alpha(E_n^{'}-E_m^{'})$. The efficiency and the PWC for such special Otto cycles are given as~\cite{Quan07}: $\eta=1-1/\alpha$ and $T_1>\alpha T_2$, respectively. Using Eq.~(\ref{eq:lmg3}) for the hot and cold heat bath cases with $J_1=rh_1$ and $J_2=rh_2$, respectively, one can calculate the proportionality constant for the energy gaps as: $\alpha=h_1/h_2$ (with $\alpha>1$). Indeed, the special coupled LMG Otto cycle efficiency and the PWC are given by those for Quan's model of special multilevel Otto cycles.

\begin{figure}[ht]
	\centering
	\resizebox{0.60\hsize}{!}{\includegraphics{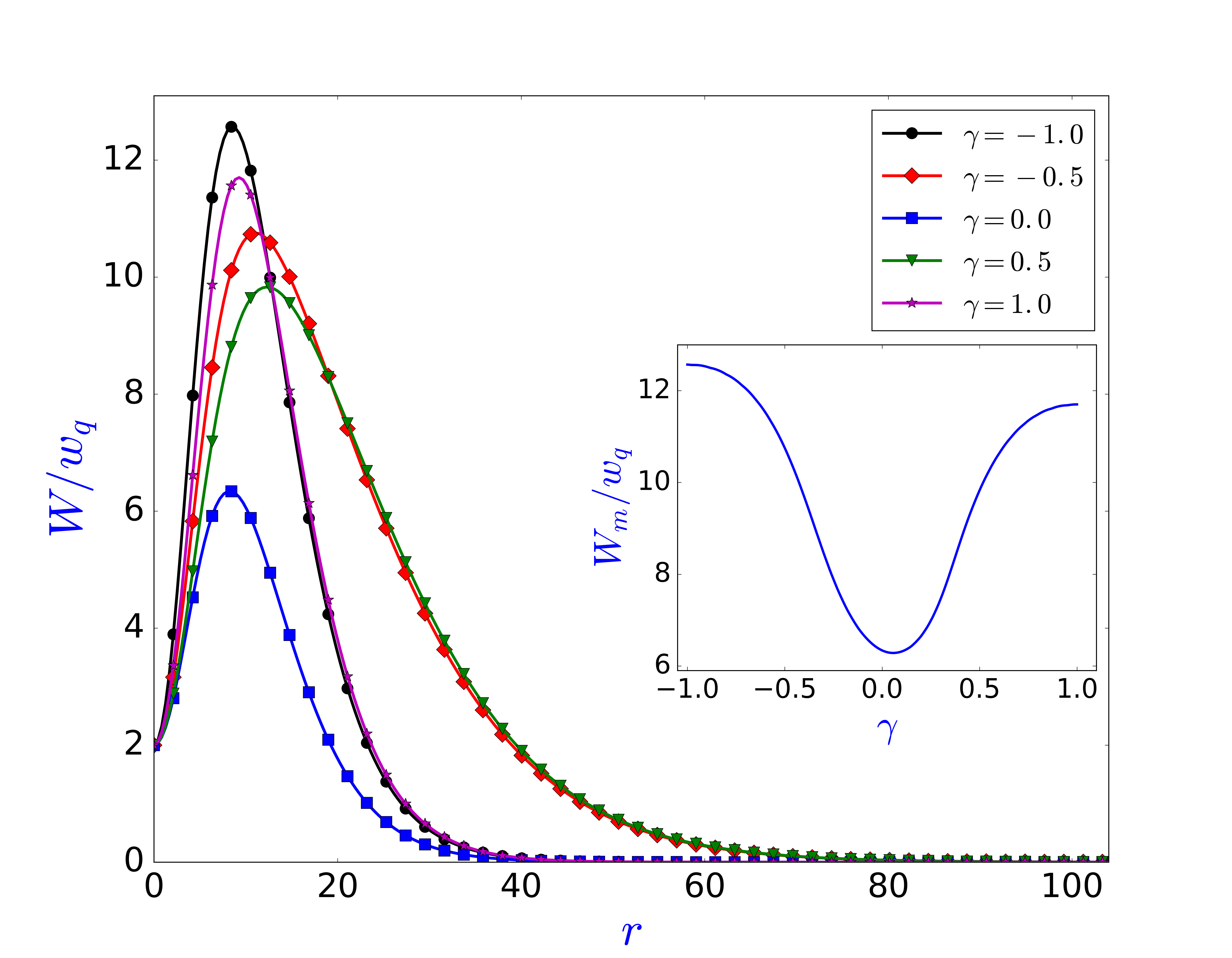}}
	\caption{(Color online.) The ratio $W/w_q$ versus relative coupling strength $r$ for the parameters  $T_1=1.0$, $T_2=0.50$, $h_1=0.50$, $h_2=0.30$, $J_1=r h_1$, $J_2=r h_2$ and $\gamma=-1.0,-0.5,0.0,0.5,1.0 $. The inset shows $\gamma$ effects on the maximum of the work output divided by $w_q$. Note that the thermal efficiency and the classical Carnot efficiency for the above parameters are $\eta=1-h_2/h_1=0.4$ and $\eta_c=1-T_2/T_1=0.5$, respectively.}	
	\label{fig:qoe6}
\end{figure}

Contrary to the previously discussed cases, the coupled system produces work under the conditions where the single qubit also acts as a quantum Otto engine. On the other hand, there is significant enhancement relatively due to cooperative work. To show the benefit of coupling on the work harvested, we plot the work output $W$ divided by the work done by a single qubit $w_q$ in the same cycle, as a function of relative coupling strength $r$ in Fig.~\ref{fig:qoe6} for the ratios $T_2/T_1=0.5$ and $h_2/h_1=0.6$. $w_q$ is obtained for the single qubit Hamiltonian $H_q=-h/2\sigma_z$ and it is fixed for the above parameters $w_q=4.6\times10^{-3}$. Furthermore, the thermodynamical equations remain invariant under the replacement $r\rightarrow -r$. The coupled engine can produce work for a wide range of relative coupling strength $r$ with a constant efficiency $\eta=0.4$ (where $\eta_c=0.5$). The regions $W/w_q>2$ indicates the coupling enhanced work output. In these regions, the coupled system can produce work more than the qubit (i.e., uncoupled system). The work output is maximum at certain $r$ and strongly depends on the anisotropy parameter $\gamma$. The dependence of $W_m$ on $\gamma$ is shown in inset of Fig.~\ref{fig:qoe6}. The ratio $W_m/w_q$ is always greater than $2$, and for the whole considered range of $\gamma$, the special coupled QHE can produce work; $\gamma=-1$ is the optimal value for the performance of the special coupled heat engine, where the ratio $W_m/w_q$ can exceed 12.

\section{\label{sec:Conclusions}Conclusions}
We have considered LMG model as the working substance of a quasi-stationary quantum Otto cycle. The adiabatic branches of the cycle are considered to involve the change of either the external magnetic field $h$ or the coupling strength $J$, or the simultaneous change of $h$ and $J$ in which the ratio $r=J/h$ is fixed during the cycle. For each cases, the anisotropy effects on the engine efficiency and work output are investigated in detail. It is found that for the adiabatic changes by either $J$ or $h$, the LMG Otto engine can harvest work under the conditions where Kieu's conditions are no longer valid. We develop the engine operations $h_1>h_2$ and $h_1<h_2$ for the $h$ changes, and $J_1>J_2$ and $J_1<J_2$ for the $J$ changes in adiabatic stages.  The LMG Otto engine is found to harvest work with an appreciably high efficiency near or comparable to the classical Carnot efficiency. For the simultaneous change of $J$ and $h$, the operation conditions of the coupled LMG model are found to be determined by the Kieu's conditions where both the coupled LMG and a qubit as an Otto cycle produce work with the same efficiency. On the other hand, cooperative nature of the system is revealed by showing that strongly anisotropic, two-axis twisting, spin interactions enhance the work output up to the twelve times of the work by the single qubit Otto engine.   Our results indicate that anisotropy is a critical parameter to consider in interacting spin QHEs to optimize cooperative work enhancement, to open new operation windows and to optimize maximum work at high efficiency.  

\section*{\label{sec:Acknowledgements}Acknowledgements}

S.\c{C}. warmly thanks A. Gen\c{c}ten, A.\"{U}.C. Hardal and D. T\"{u}rkpen\c{c}e for useful discussions. F.A. acknowledges the support and the hospitality of the office of Vice President for Research and Development (VPRD) and Department of Physics of the Ko\c{c} University. The authors acknowledge support from Ko\c{c} University and Lockheed Martin Corporation Research Agreement.


\begin{thebibliography}{}

\bibitem{scovil59} H.E.D. Scovil, E.O. Schulz-Dubois, Phys. Rev. Lett. \textbf{2}, 262 (1959)
\bibitem{Quan07} H.T. Quan, Y.-X Liu, C.P. Sun, F. Nori, Phys. Rev. E \textbf{76}, 031105 (2007)  
\bibitem{Quan09} H.T. Quan, Phys. Rev. E \textbf{79}, 041129 (2009) 
\bibitem{Kieu04} T.D. Kieu, Phys. Rev. Lett. \textbf{93}, 140403 (2004)
\bibitem{Kieu06} T.D. Kieu, Eur. Phys. J. D \textbf{39}, 115-128 (2006)
\bibitem{Dillen09} R. Dillenschneider, E. Lutz, EPL \textbf{88} (2009) 50003 
\bibitem{Scully03} M.O. Scully, M.S. Zubairy, G.S. Agarwal, H. Walther, Science \textbf{299}, 862 (2003) 
\bibitem{Hardal15} A.\"{U}.C. Hardal, \"{O}.E. M\"{u}stecapl{\i}o\u{g}lu, Sci. Rep. \textbf{5}, 12953 (2015)
\bibitem{Huang14} X.L. Huang, X.Y. Niu, X.M. Xiu, X.X. Yi, Eur. Phys. J. D \textbf{68}, 32 (2014)	
\bibitem{Quan05} H.T. Quan, P. Zhang, C.P. Sun, Phys. Rev. E \textbf{72}, 056110 (2005) 	
\bibitem{Lin03} B.H. Lin, J.C. Chen, Phys. Rev. E \textbf{67}, 046105 (2003)
\bibitem{altintas14} F. Altintas, A.\"{U}.C. Hardal, \"{O}.E. M\"{u}stecapl{\i}o\u{g}lu, Phys. Rev. E \textbf{90}, 032102 (2014)
\bibitem{thomas11}  G. Thomas, R.S. Johal, Phys. Rev. E \textbf{83}, 031135 (2011)
\bibitem{zhang08} G.-F. Zhang, Eur. Phys. J. D \textbf{49}, 123-128 (2008)
\bibitem{huang13} X.L. Huang, L.C. Wang, X.X. Yi, Phys. Rev. E \textbf{87}, 012144 (2013)
\bibitem{feldmann04} T. Feldmann, R. Kosloff, Phys. Rev. E \textbf{70}, 046110 (2004)
\bibitem{feldmann03} T. Feldmann, R. Kosloff, Phys. Rev. E \textbf{68}, 016101 (2003)
\bibitem{huang213} X.L. Huang, H. Xu, X.Y. Niu, Y.D. Fu, Phys. Scr. \textbf{88}, 065008 (2013)
\bibitem{henrich07} M.J. Henrich, G. Mahler, M. Michel, Phys. Rev. E \textbf{75}, 051118 (2007)
\bibitem{zhang07} T. Zhang, W.-T. Liu, P.-X. Chen, C.-Z. Li, Phys. Rev. A \textbf{75}, 062102 (2007)
\bibitem{thomas14} G. Thomas, R.S. Johal, Eur. Phys. J. B  \textbf{87}, 166 (2014)
\bibitem{huang12} X.L. Huang, T. Wang, X.X. Yi, Phys. Rev. E \textbf{86}, 051105 (2012)
\bibitem{wu06} F. Wu, L. Chen, F. Sun, C. Wu, Q. Li, Phys. Rev. E \textbf{73}, 016103 (2006)
\bibitem{ivanchenko15} E.A. Ivanchenko, Phys. Rev. E \textbf{92}, 032124 (2015)
\bibitem{wang12} H. Wang, G. Wu, D. Chen, Phys. Scr. \textbf{86}, 015001 (2012)
\bibitem{he12} X. He, J. He, J. Zheng, Physica A \textbf{391}, 6594 (2012)
\bibitem{huang14plus} X.L. Huang, Y. Liu, Z. Wang, X.Y. Niu, Eur. Phys. J. Plus \textbf{129}, 4 (2014)
\bibitem{wang09} H. Wang, S. Liu, J. He, Phys. Rev. E \textbf{79}, 041113 (2009)
\bibitem{hübner14} W. Hubner, G. Lefkidis, C.D. Dong, D. Chaudhuri, L. Chotorlishvili, J. Berakdar, Phys. Rev. B \textbf{90}, 024401 (2014)
\bibitem{azimi14} M. Azimi, L. Chotorlishvili, S.K. Mishra, T. Vekua, W. Hubner, J. Berakdar,  New J. Phys. \textbf{16}, 063018 (2014)
\bibitem{albayrak13} E. Albayrak, Int. J. Quantum Inform. \textbf{11}, 1350021 (2013)
\bibitem{he2012} J.-Z. He, X. He, J. Zheng, Int. J. Theor. Phys. \textbf{51}, 2066 (2012)
\bibitem{alti215} F. Altintas, \"{O}.E. M\"{u}stecapl{\i}o\u{g}lu, Phys. Rev. E \textbf{92}, 022142 (2015)
\bibitem{Lutz14} J. Roßnagel, O. Abah, F. Schmidt-Kaler, K. Singer, E. Lutz, Phys. Rev. Lett. \textbf{112}, 030602 (2014)	
\bibitem{Lutz12} O. Abah, J. Roßnagel, G. Jacob, S. Deffner, F. Schmidt-Kaler, K. Singer, E. Lutz, Phys. Rev. Lett. \textbf{109}, 203006   (2012)	
\bibitem{Fialko12} O. Fialko, D.W. Hallwood, Phys. Rev. Lett. \textbf{108}, 085303 (2012)		
\bibitem{Zhang14} K. Zhang, F. Bariani, P. Meystre, Phys. Rev. Lett. \textbf{112}, 150602 (2014)			
\bibitem{Sothmann12} B. Sothmann, M. B\"{u}ttiker, EPL \textbf{99}, 27001 (2012)		
\bibitem{Quan06} H.T. Quan, P. Zhang, C.P. Sun, Phys. Rev. E \textbf{73}, 036122 (2006)	
\bibitem{Altintas15} F. Altintas, A.\"{U}.C. Hardal, \"{O}.E. M\"{u}stecapl{\i}o\u{g}lu, Phys. Rev. A \textbf{91}, 023816 (2015)	
\bibitem{Lipkin1965} H.J. Lipkin, N. Meshkov, A.J. Glick, Nucl. Phys. \textbf{62}, 188 (1965)
\bibitem{Meshkov1965} N. Meshkov, A.J. Glick, H.J. Lipkin, Nucl. Phys. \textbf{62}, 199 (1965)	
\bibitem{Glick1965} A.J. Glick, H.J. Lipkin, N. Meshkov, Nucl. Phys. \textbf{62}, 211 (1965)
\bibitem{Heggie1998} M.I. Heggie, M. Terrones, B.R. Eggen, G. Jungnickel, R. Jones, C.D. Latham, P.R. Briddon, H. Terrones, Phys. Rev. B \textbf{57}, 13339 (1998)
\bibitem{Chen09} G. Chen, J.-Q. Liang, S. Jia, Optics Express \textbf{17}, 19682 (2009)	
\bibitem{Cirac1998} J.I. Cirac, M. Lewenstein, K. Molmer, P. Zoller, Phys. Rev. A  \textbf{57}, 1208 (1998)	
\bibitem{Morrison2008a} S. Morrison, A.S. Parkins, Phys. Rev. A \textbf{77}, 043810 (2008)
\bibitem{Morrison2008b} S. Morrison, A.S. Parkins, Phys. Rev. Lett. \textbf{100}, 040403 (2008)
\bibitem{Hamdouni2007} Y. Hamdouni, F. Petruccione, Phys. Rev. B \textbf{76}, 174306 (2007)
\bibitem{Quan2007a}	H.T. Quan, Z.D. Wang, C.P. Sun, Phys. Rev. A \textbf{76}, 012104 (2007)
\bibitem{Das2006} A. Das, K. Sengupta, D. Sen, B.K. Chakrabarti, Phys. Rev. B \textbf{74}, 144423 (2006)
\bibitem{vidal04} J. Vidal, G. Palacios, G. Aslangul, Phys. Rev. A \textbf{79}, 062304 (2004)
\bibitem{ma2009} J. Ma, X. Wang, Phys. Rev. A \textbf{80}, 012318 (2009)
\bibitem{ma2011} J. Ma, X. Wang, C.P. Sun, F. Nori, Physics Reports \textbf{509}, 89 (2011)

\bibitem{vidal06} J. Vidal, Phys. Rev. A \textbf{73}, 062318 (2006)
\bibitem{vidal12} J. Wilms, J. Vidal, F. Verstraete, S. Dusuel, J. Stat. Mech. \textbf{P01023}, (2012)
\bibitem{vidal10} H. Wichterich, J. Vidal, S. Bose, Phys. Rev. A \textbf{81}, 032311 (2010)

\bibitem{Salvatori14} G. Salvatori, A. Mandarino, M.G.A. Paris, Phys. Rev. A \textbf{90}, 022111 (2014)
\bibitem{Kitagawa93} M. Kitagawa, M. Ueda, Phys. Rev. A \textbf{47}, 5138 (1993)
\bibitem{eric2010} E. Sj\"oqvist, R. Rahaman, U. Basu, B. Basu, J. Phys. A: Math. Theor. \textbf{43}, 354026 (2010)

\bibitem{Rigolin12} G. Rigolin, G. Ortiz, Phys. Rev. A \textbf{85}, 062111 (2012)
\bibitem{Caneva08} T. Caneva, R. Fazio, G.E. Santoro, Phys. Rev. B \textbf{78}, 104426 (2008)
\bibitem{Caneva09} T. Caneva, R. Fazio, G.E. Santoro, J. Phys.: Conf. Ser. \textbf{143}, 012004 (2009)
\bibitem{Solinas08} P. Solinas, P. Ribeiro, R. Mosseri, Phys. Rev. A \textbf{78}, 052329 (2008)

\bibitem{Zheng15} Y. Zheng, S. Campbell, G. D. Chiara, D. Poletti, arXiv:1509.01882, (2015)
 
\end{thebibliography}
\end{document}